\documentclass[%
  superscriptaddress,
   amsmath,amssymb,
   aps,
   prb,twocolumn
  ]{revtex4-1}

\usepackage{graphicx}
\usepackage{xcolor}
\usepackage{bm}
\usepackage{placeins}
\usepackage{amssymb}
\usepackage{acro}
\usepackage{xspace}
\usepackage{subfig}
\usepackage{multirow}
\usepackage[percent]{overpic}

\DeclareAcronym{BZ}{
  short=BZ,
  long=Brillouin zone,
  }
\DeclareAcronym{TWS}{
  short=TWS,
  long=topological Weyl semimetal
  }
\DeclareAcronym{WS}{
  short=WS,
  long=Weyl semimetal
  }
\DeclareAcronym{TDS}{
  short=TDS,
  long=topological Dirac semimetal
  }
\DeclareAcronym{WN}{
  short=WN,
  long=Weyl node
  }
\DeclareAcronym{3D}{
  short=3D,
  long=three-dimensional,
  }
\DeclareAcronym{2D}{
  short=2D,
  long=two-dimensional,
  }
\DeclareAcronym{SOI}{
  short=SOI,
  long=spin-orbit interaction,
  }
\DeclareAcronym{SOC}{
  short=SOC,
  long=spin-orbit coupling,
  }
\DeclareAcronym{SO}{
  short=SO,
  long=spin-orbit,
  }
\DeclareAcronym{ARPES}{
  short=ARPES,
  long=angular resolved photoemission spectroscopy,
  }
\DeclareAcronym{DFT}{
  short=DFT,
  long=density-functional theory,
  }
\DeclareAcronym{XC}{
  short=XC,
  long=exchange and correlation,
  }
\DeclareAcronym{PBE}{
  short=PBE,
  long={Perdew, Burke and Ernzerhof},
  }
\DeclareAcronym{GGA}{
  short=GGA,
  long=generalized gradient approximation,
  }
\DeclareAcronym{PW}{
  short=PW,
  long=plane wave,
  }
\DeclareAcronym{EELS}{
  short=EELS,
  long=electron energy loss spectra,
  }
\DeclareAcronym{bct}{
  short=bct,
  long=body-centered tetragonal,
  }
\DeclareAcronym{QE}{
  short=QE,
  long=\texttt{Quantum ESPRESSO},
  }
\DeclareAcronym{DOS}{
  short=DOS,
  long=density of states,
  }
\DeclareAcronym{IR}{
  short=IR,
  long=infrared,
  }
\DeclareAcronym{UV}{
  short=UV,
  long=ultraviolet,
  }
\DeclareAcronym{TaAs}{
  short=TaAs,
  long=tantalum arsenide,
  }
\DeclareAcronym{scf}{
  short=scf,
  long=self-consistent,
  }
\DeclareAcronym{HWCC}{
  short=HWCC,
  long=hybrid Wannier charge centers,
  }
\DeclareAcronym{QSHI}{
  short=QSHI,
  long=quantum spin Hall insulator,
  }
\DeclareAcronym{QSHE}{
  short=QSHE,
  long=quantum spin Hall effect,
  }
\DeclareAcronym{TR}{
  short=TR,
  long=time-reversal,
  }
\DeclareAcronym{TRIM}{
  short=TRIM,
  long=time-reversal invariant momenta,
  }
\DeclareAcronym{vdW}{
  short=vdW,
  long=van-der-Waals,
  }
\DeclareAcronym{DG}{
  short=DG,
  long=direct gap,
  }
\DeclareAcronym{IG}{
  short=IG,
  long=indirect gap,
  }
\DeclareAcronym{DGM}{
  short=DGM,
  long=direct gap metal,
  }

\newcommand{\angstrom}{\textup{\AA}}

\newcommand{\zz}{$\mathbb{Z}_2$\xspace}

\begin{document}

  \title{Complementary screening for quantum spin Hall insulators in two-dimensional exfoliable materials}

  \author{Davide Grassano}
  \email{davide.grassano@epfl.ch}
  \affiliation{Theory and Simulations of Materials (THEOS) and National Center for Computational Design and Discovery of Novel Materials (MARVEL), \'Ecole Polytechnique F\'ed\'erale de Lausanne, CH-1015 Lausanne, Switzerland}

  \author{Davide Campi}
  \affiliation{Theory and Simulations of Materials (THEOS) and National Center for Computational Design and Discovery of Novel Materials (MARVEL), \'Ecole Polytechnique F\'ed\'erale de Lausanne, CH-1015 Lausanne, Switzerland}
  \affiliation{Dipartimento di Scienza dei Materiali, Universita di Milano-Bicocca, Via Cozzi 53, 20125 Milano, Italy}
  
  \author{Antimo Marrazzo}
  \affiliation{Theory and Simulations of Materials (THEOS) and National Center for Computational Design and Discovery of Novel Materials (MARVEL), \'Ecole Polytechnique F\'ed\'erale de Lausanne, CH-1015 Lausanne, Switzerland}
  \affiliation{\trieste}

  \author{Nicola Marzari}
  \affiliation{Theory and Simulations of Materials (THEOS) and National Center for Computational Design and Discovery of Novel Materials (MARVEL), \'Ecole Polytechnique F\'ed\'erale de Lausanne, CH-1015 Lausanne, Switzerland}

  \newcommand{\trieste}{Dipartimento di Fisica, Universit\`a di Trieste,  I-34151 Trieste, Italy}

  \begin{abstract}
    \Aclp{QSHI} are a class of topological materials that has been extensively studied during the past decade.
    One of their distinctive features is the presence of a finite band gap in the bulk and gapless, topologically protected edge states that are spin-momentum locked.
    These materials are characterized by a \zz topological order where, in the 2D case, a single topological invariant can be even or odd for a trivial or a topological material, respectively.
    Thanks to their interesting properties, such as the realization of dissipationless spin currents, spin pumping and spin filtering, they are of great interest in the field of electronics, spintronics and quantum computing.
    In this work we perform an high-throughput screening of \aclp{QSHI} starting from a set of 783 \acl{2D} exfoliable materials, recently identified from a systematic screening of the ICSD, COD, and MPDS databases.
    We find a new \zz topological insulator (HgNS) as well as 3 already known ones and 7 \aclp{DGM} that have the potential of becoming \aclp{QSHI} under a reasonably weak external perturbation.
    
  \end{abstract}

  \maketitle

  \acresetall
  \section{Introduction}
    Topological states of matter\cite{wang2017topological} have been one of the main subject of studies for condensed matter physicists in the past decade.
    One of the driving aspects lies in the robustness of many properties of topological materials, whose existence  can be directly related to a specific set of topological invariants\cite{niu1985quantized,fu2006time,soluyanov2012thesis}.
    In the case of insulators, the ground state of a topologically trivial material can be connected to the one of an atomic insulator or vacuum by a smooth adiabatic deformation of its Hamiltonian, while for a topologically non-trivial material, this cannot happen without closing a gap in its band structure.
    Among the possible classifications of topological insulators, \acp{QSHI} represent a class of materials protected by \ac{TR} symmetry where the \ac{QSHE}\cite{kane2005z} can be realized.
    This is made possible by the presence of a finite bulk gap and spin-polarized gapless helical hedge states\cite{dai2008helical} that can sustain dissipationless spin currents\cite{wang2017topological} thanks to the absence of backscattering forbidden by \ac{TR}. 
    Other interesting properties arise from spin-momentum locking\cite{timm2012transport}, which allows for the possibility to realize spin filtering\cite{rachel2014giant}, injection, detection\cite{abanin2007charge} and pumping\cite{fu2006time}, all of which make \acp{QSHI} of great interest for the realization of spintronic devices.
    It has also been shown that magnetic confinement can be used in order to realize qubits on the surface of a \ac{QSHI}\cite{ferreira2013magnetically}, making these materials promising for applications in quantum computing.
    The need for finding new topological materials, and in particular \acp{QSHI}, is further accentuated by the recent surge in high-throughput studies\cite{li2018high,marrazzo2019relative,olsen2019discovering,vergniory2019complete,tang2019comprehensive,zhang2019catalogue,gjerding2021recent}, especially when related to easily exfoliable materials\cite{mounet2018two,campi2023expansion,shipunov2021layered,guo2020quantum}. 

    \acp{QSHI} can be identified through a \zz invariant which can be either even (0) or odd (1), as first proposed by Kane and Mele, that showed that a Dirac cone, with a gap opened by \ac{SOC} can lead to a topologically non-trivial order\cite{kane2005z,kane2005quantum}.
    This concept has later been extended as, in general, a \ac{QSHI} can be obtained whenever a band inversion takes place, as in the paradigmatic cases of, e.g., HgTe quantum dots\cite{bernevig2006quantum,konig2007quantum} and Bi$_{1-x}$Sb$_x$\cite{fu2007topological}.
    A non-trivial \zz invariant can emerge naturally due to the material intrinsic band structure, or can be induced by a perturbation, such as the application of strain or electric fields or the interaction with a substrate \cite{giovannetti2007substrate,Drummond.2012,grassano2018detection,collins2018electric,zhao2020strain}.
    Hence, a trivial insulator can be driven into a topological insulator and vice-versa, leaving open the possibility to realize a topological transistor\cite{liu2014spin,qian2014quantum}.
    While for \ac{2D} materials a single invariant is sufficient in order to describe their topology with respect to the \ac{QSHE}, it has been shown that for \ac{3D} materials a set of 4 invariants is required, leading to the possibility of weak or strong \zz topological insulators\cite{fu2007threedim}.
    It has also been shown that the topology of a system can often be related to its symmetries\cite{slager2013space,kruthoff2017topological,po2017symmetry,khalaf2018symmetry,song2018quantitative}.
    For example, in the case of \acp{QSHI} with inversion symmetry, the \zz topological invariant can be computed easily by calculating the eigenvalues of the parity operator at the \ac{TRIM} points of the \ac{BZ}\cite{fu2007topological}.

    In this work we perform a complementary screening to that of Marrazzo et al. \cite{marrazzo2019relative}, by using a set of newly discovered \ac{2D} materials from Ref \onlinecite{MC2D}, 
    in which 1252 new monolayers, exfoliable from their layered (typically \acl{vdW} bonded) parents, have been added to the initial \ac{2D} portfolio\cite{mounet2018two} used in Ref \onlinecite{marrazzo2019relative}.
    We set out with this, the aim of identifying all the \ac{2D} \zz topological insulators present in this set, analyzing all the materials with less than 40 atoms per unit cell, with the exception of lanthanides due to their typically complex, correlated electronic structure\cite{gillen2013nature}.
    We also identify \acp{DGM} which can serve as \zz candidates, by calculating the invariant for materials where the valence and conduction band manifolds are separated by a \ac{DG}, but where the \ac{IG} is negative.

  \section{Methods}
    All the \ac{DFT} calculations performed in this work use the \ac{QE}\cite{Giannozzi.Baroni.ea:2009:JoPCM,espresso2017} distribution.
    We perform fully relativistic calculations with \ac{SOC}, using the \ac{PBE} pseudopotentials\cite{perdew1996generalized} from the ONCVPSP (v0.4) set\cite{PhysRevB.88.085117} contained in the PseudoDojo library\cite{VANSETTEN201839}.
    For every calculations we use a kinetic cutoff, high enough to ensure that the uncertainty on the energy eigenvalues is lower than 5~meV.
    In order to properly account for low-gap and metallic systems, a Fermi-Dirac smearing of 5~meV is used.
    For the initial self consistent calculations we use a {\bf k}-point mesh of density $0.2$~\AA$^{-1}$, which is refined to $0.1$~\AA$^{-1}$ for \ac{QSHI} candidates.
    All the calculations have been performed assuming a non-magnetic ground state, magnetism has been later checked for the selected candidates following the same approach used in Ref.\onlinecite{mounet2018two}.
    
    The topological invariant are computed either from an analysis of the band structures at the \ac{TRIM} points or using the Z2Pack package\cite{gresch2017z2pack}, for which we set the maximum number of k-points per line to 200, and the lowest acceptable distance between lines to $10^{-7}$.
    The threshold on the \ac{HWCC}\cite{sgiarovello2001electron} position variation per step on a line (pos\_tol) is set to 0.01, the threshold on the relative movement of the \acp{HWCC} position between neighboring (move\_tol) to 0.3 and the threshold for the distance between the middle of the largest gap in a line and the position of the neighboring \ac{HWCC} (gap\_tol) to 0.3.

    For the new \ac{QSHI} found, we also perform a hybrid functional calculation, using HSE06\cite{heyd2003hybrid}.
    We find that in order to obtain a properly converged Wannierization of the system, a very dense $32 \times 16 \times 1$ k-point mesh with a $4 \times 2 \times 1$ q-point mesh has to be used.
    The Wannierized wavefunctions are then used to compute the topological invariant and edge states, in order to confirm whether the system remains a topological insulator.

    \subsection{Screening procedure}
      The original set of material of which we explore the properties is derived from the ICSD\cite{bergerhoff1987crystallographic,zagorac2019recent}, COD\cite{COD} and MPDS\cite{villars2004pauling} databases, by identifying \ac{2D} exfoliable materials, starting from experimentally known bulk structures using \ac{vdW} DF2-C09 \ac{DFT} calculations\cite{mounet2018two}.
      First we exclude all materials containing lanthanides, due to the inaccuracy of plain \ac{DFT} pertaining calculations with these atoms caused by the presence of strongly correlated electrons\cite{gillen2013nature}.
      For the remaining 603 materials, compared to the 1306 structures studied in Ref \onlinecite{marrazzo2019relative} by Marrazzo et al, we follow two strategies for the identification of \acp{QSHI} in order to minimize the usage of computational resources.
      For the ones with inversion symmetry we use the formula given by Fu and Kane\cite{fu2007topological}
      \begin{equation}\label{eq:fu_kane_z2_inversion}
        (-1)^\nu
        =
        \prod_i
          \delta _i
        , \quad \quad
        \delta _i
        =
        \prod_{m=1}^N
          \xi _{2 m} (\Gamma _i)
      \end{equation}
      where $\nu$ is the \zz topological invariant, $\xi _{2 m}$ is the eigenvalue of the parity operator calculated for a couple of \ac{TR}-paired bands and $\Gamma _i$ are the \ac{TRIM} points, of which there are 4 for a \ac{2D} material, defined as
      \begin{equation}
        \Gamma _i = - \Gamma _i + {\mathbf G}
      \end{equation}
      with ${\mathbf G}$ being any reciprocal lattice vector.
      For a \ac{2D} material, 4 \ac{TRIM} points can be found at the coordinates $(0, 0)$, $(0, 0.5)$, $(0.5, 0)$ and $(0.5, 0.5)$ expressed in units of the reciprocal lattice vector.
      Given the formula in eq.\eqref{eq:fu_kane_z2_inversion}, the most computationally efficient route for materials that have inversion symmetry is to calculate first the band structure at the \ac{TRIM} points to determine the topological invariants.
      Materials with $\nu = 0$ are discarded, while for those with $\nu = 1$ we proceed with a band structure calculation along the high-symmetry lines.
      The result can either be a material with finite \ac{DG} on the entire {\bf k}-point path, or with a zero \ac{DG}.
      The latter is discarded as a metal/semimetal without an isolated valence band manifold, while the former can either be a \ac{QSHI} if the \ac{IG}, defined as the difference between the minimum of the conduction band and the maximum of the valence band, is positive, or a \ac{DGM} if the \ac{IG} is negative.
      The values of the \acp{DG} and \acp{IG} are further corroborated by analyzing the band structure computed on a dense uniform k-point grid with a spacing of 0.015~\angstrom$^-1$.
      We discard materials that are \ac{DGM} with an \ac{IG} lower than -0.15~eV, as attempting to drive such a material to a topological phase would be hardly feasible.
      The same band structure on the dense uniform grid is also used to derive the \ac{DOS} for the materials studied.

      If inversion symmetry is not present, we follow a different route, by determining the \ac{DG} and \ac{IG} first through a band structure calculation and discarding the metals and semimetals without a finite \ac{DG}.
      We then use the Z2Pack package, which tracks the positions of one-dimensional \acp{HWCC} across half of the \ac{BZ}, in order to compute the topological invariant\cite{soluyanov2011computing}.
      The package acts as an automation tool, which interfaces a \ac{DFT} code (\ac{QE} in our case) with Wannier90\cite{pizzi2020wannier90} in order to calculate the \ac{HWCC} positions on a progressively denser grid of {\bf k}-points.
      If one of the convergence criteria of Z2Pack, with the parameters mentioned before, is not satisfied, then the material is discarded.
      If the results of the calculation is a non trivial topological invariant, the material is then classified as a \ac{QSHI} or \ac{DGM} following the same criteria mentioned for the inversion-symmetric cases, where the direct and indirect gaps are then checked again from the band structures on a dense k-point grid used also for the \ac{DOS}.
      
\begin{figure}
        \includegraphics[width=0.4\textwidth, trim=0 0 0 0]{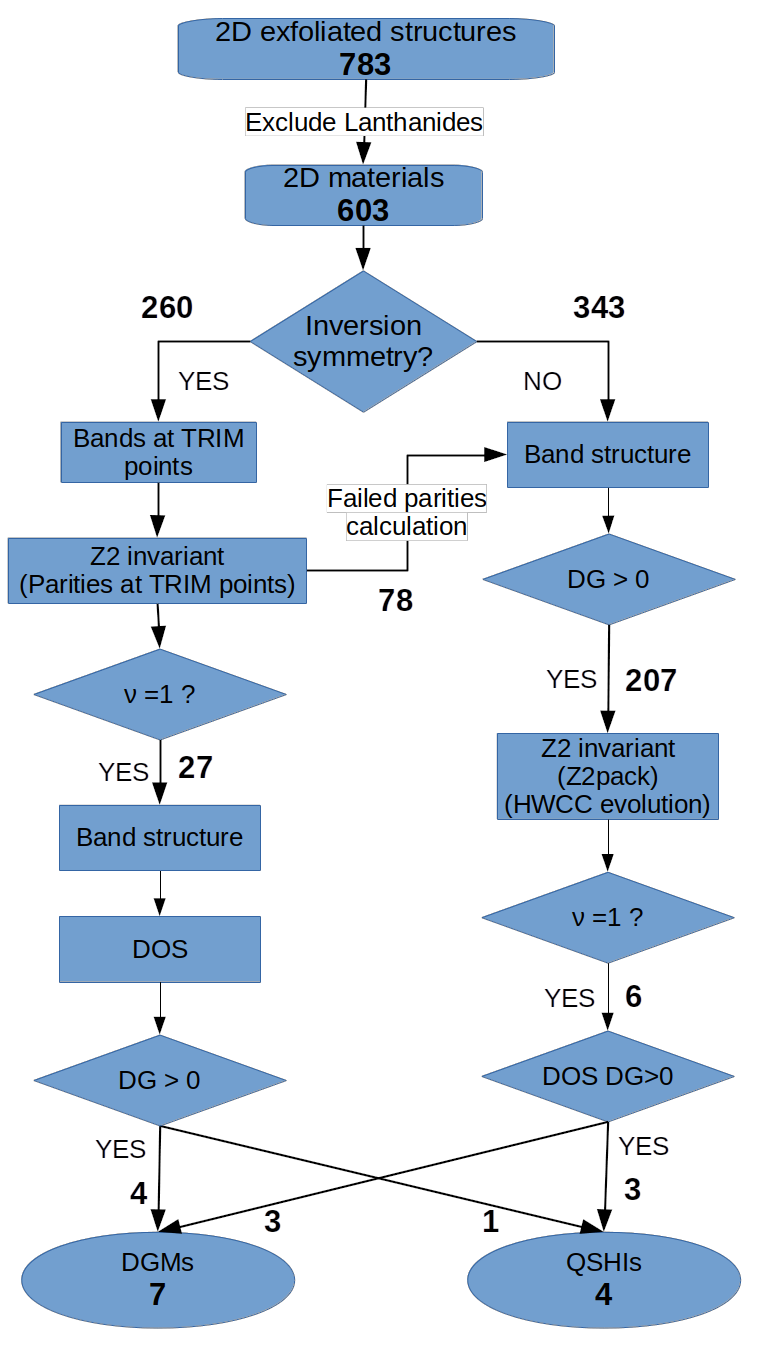}
        \caption{
          \label{fig:flowchart}
          Workflow adopted for the high-throughput screening of the QSHIs.
          Starting from 783 structures presented in Ref \onlinecite{MC2D} we proceed to a study of the \zz topological invariant following two branches, depending on the presence of inversion symmetry in the material.
          The steps in both branches are organized as to limit the usage of resources, by performing the lowest cost screenings first in each branch.
        }
      \end{figure}

      Finally, for the selected candidates we asses the dynamical stability by computing the full phonon dispersion by means of density-functional perturbation theory\cite{baroni2001phonons} and we check the non-magnetic ground state assumption comparing the energy of the non-magnetic ground state with different possible ferromagnetic, antiferromagnetic and ferrimagnetic configurations following the same procedure adopted in Ref.~\onlinecite{mounet2018two}.
      A flowchart describing the entire high-throughput calculation is shown in Fig.~\ref{fig:flowchart}.
      The entire process has been automated using AiiDA\cite{pizzi2016aiida,huber2020aiida}, a workflow managing infrastructure, which enable us to keep track of the provenance at every computational step.
      The data produced in this work, including both the results and the entire provenance tree, has been made available on the Materials Cloud as Ref.~\onlinecite{grassano2022mclink}.

  \section{Results and discussions}
    We start our screening from a set of 783 \ac{2D} exfoliable structures taken from Ref \onlinecite{MC2D}, which are reduced to 603 after removing materials containing lanthanides.
    The set is further divided in two groups of 260/342 structures with/without inversion symmetry.
    For the inversion symmetric materials, we compute the eigenvalues of the parity operator $\xi _{2m}$, the calculation of which can sometimes fail due to limitations in the code to handle non-symmorphic space groups, or due to the accuracy threshold imposed for the calculation of the trace of the representation for each group of bands.
    78 out of these 260 \ac{TRIM} point calculation failed and were hence recalculated using Z2pack.
    Of the 182 successful calculations, 27 resulted in a non trivial topological invariant $\nu = 1$.
    We then performed a band structure calculations along the high-symmetry path\cite{hinuma2017band}, from which we excluded 22 metallic materials and identified 1 \ac{QSHI} (HfBr) and 4 \acfp{DGM}.

    On the 343+78 remaining structures (the non-inversion symmetric ones and those with failed parity calculations), the band structure calculations show that 207 materials have a \ac{DG} greater than 0; for them we proceed with Z2Pack calculations that identify, 3 \acp{QSHI} (ZrTe$_{5}$, HgNS, HfTe$_{5}$) and 3 \acp{DGM}.
    The summary of the data for every material is reported in Table~\ref{table:results}, while a more detailed report is given in the supplementary materials\cite{supplementary}, together with the crystal structure and band structure of the material identified and a plot of the \ac{HWCC} for every Z2Pack calculation.

\begin{figure*}
      \includegraphics[width=0.9\textwidth, trim=0 0 0 0]{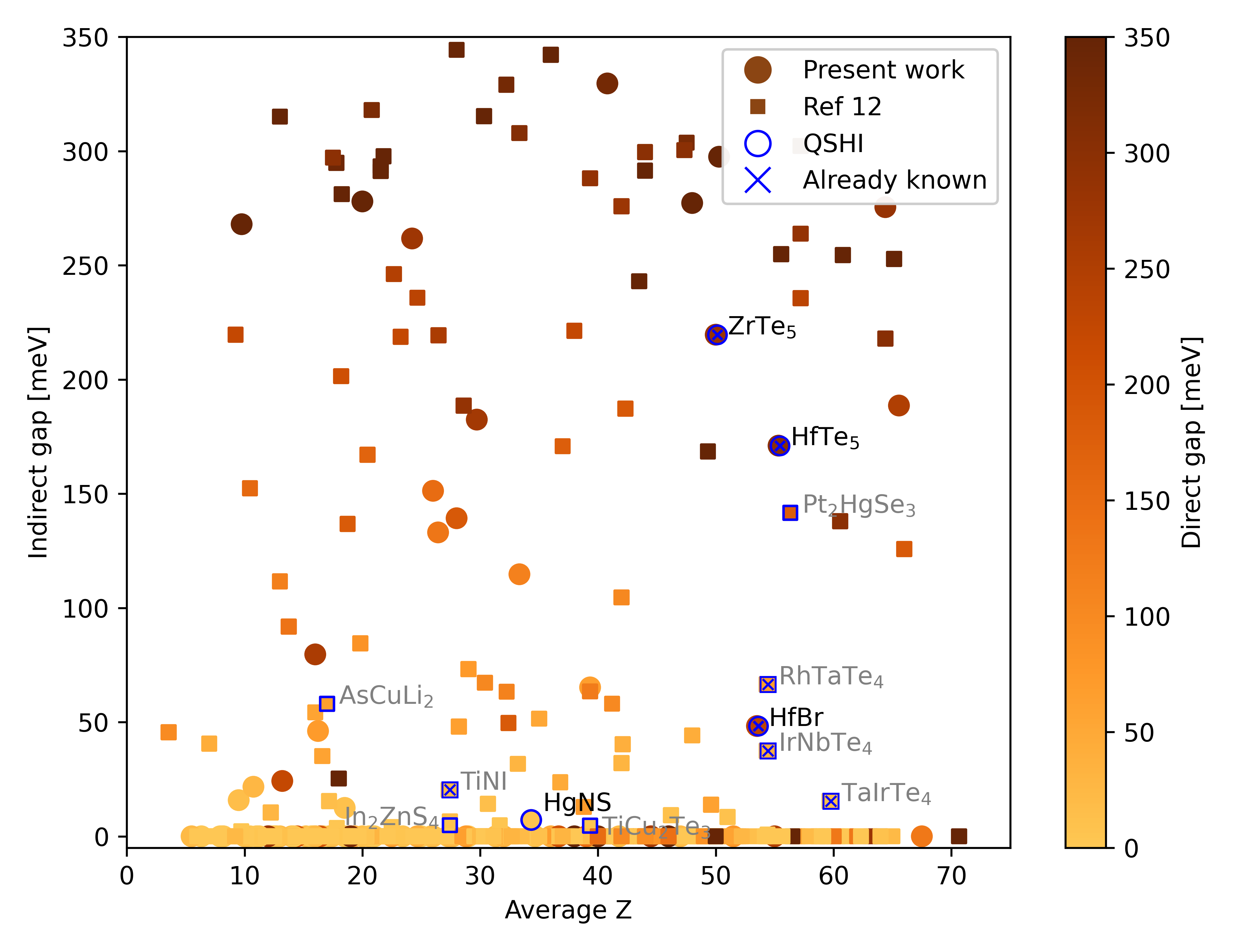}
      \caption{
        \label{fig:band_gaps}
        Plot of the direct (color scale) and indirect (y-axis) band gaps in the presence of \ac{SOC} for all the material screened in the current work (circles) and in the previous work of Marrazzo et al.\cite{marrazzo2019relative} (squares).
        Materials with negative \acp{IG} have been given a zero value.
        The \ac{DG} for each material is given using the color scale on the right.
        \Acp{QSHI} identified by the present screening effort and that of Ref \onlinecite{marrazzo2019relative} are highlighted using a blue contour to the circle/squares.
        Materials already known in the literature are also highlighted with a cross mark.
      }
    \end{figure*}

    A summary for all the materials analyzed in the present work, together with those of the previous screening by Marrazzo et al.\cite{marrazzo2019relative} (in absence of strain), is shown in Fig.~\ref{fig:band_gaps}, where the \acp{QSHI} are highlighted.
    In the figure, all materials with a negative \ac{IG} have been put on the $y=0$ axis in order to emphasize \acp{QSHI} and \acp{DGM}, especially with regard to the magnitude of the \ac{DG} and \ac{IG}, as these two parameters can be used to determine the range of temperatures at which a device could operate\cite{gusev2014temperature}.

\begin{figure}
      \includegraphics[width=0.4\textwidth, trim=0 0 0 0]{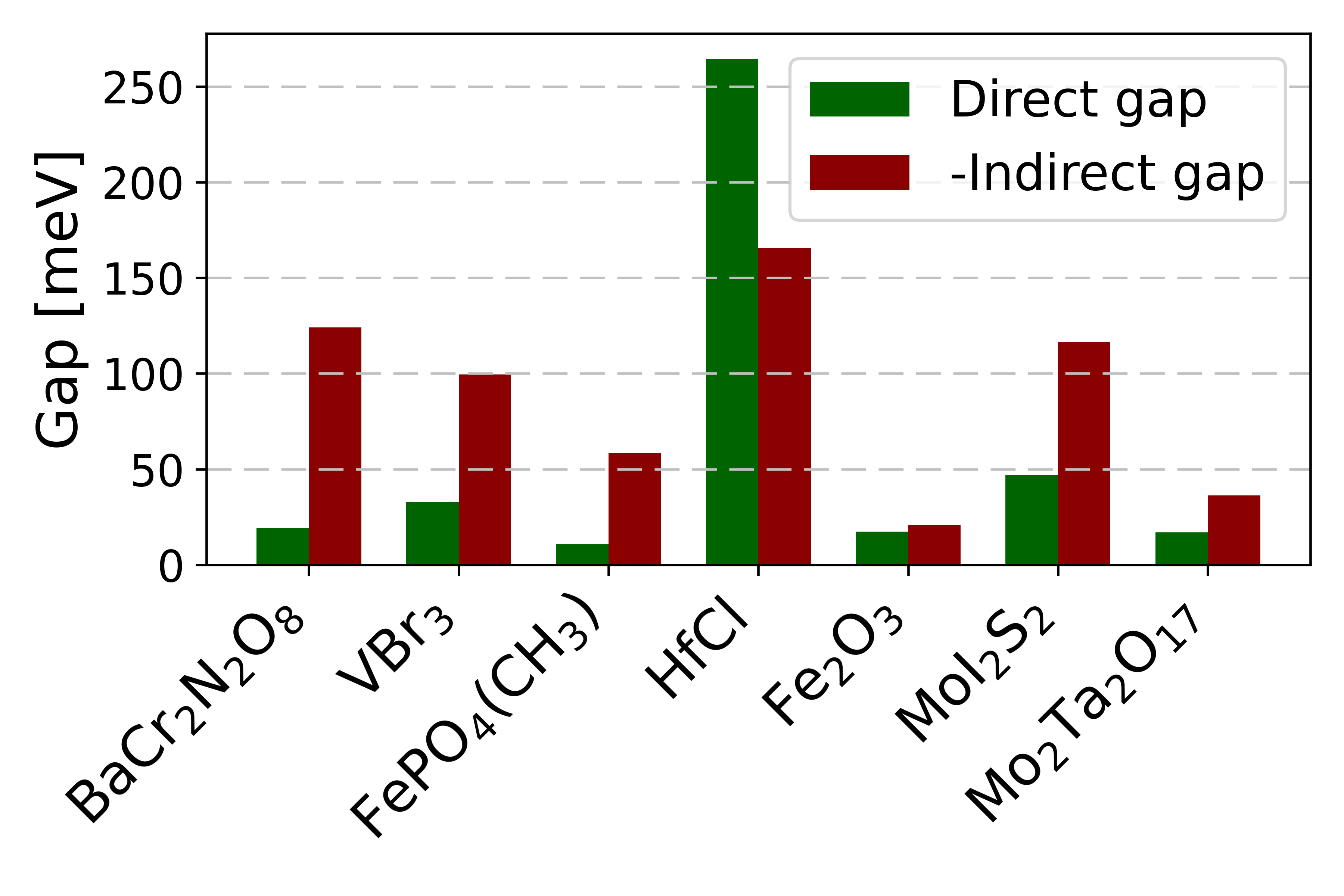}
      \caption{
        \label{fig:dmg_hyst}
        Plot of the direct and indirect  band gaps of the \acp{DGM} identified.
        The criteria used are that the material has to exhibit \ac{DG} $>$ 0 (which allows for the calculation of \zz on an isolated manifold) and a negative \ac{IG} $<$ 300~meV.
        These materials would become \acp{QSHI} in the presence of perturbations such as strain, substrate interactions or an electric field capable of opening the \ac{IG} without causing a band inversion.
      }
    \end{figure}

    The interest in \acp{DGM} with a non trivial $\nu$ is twofold.
    First it is possible that, given the limitations of DFT-PBE calculations, the band gap of the material is being underestimated and, using more appropriate methods (such as many-body perturbation theory in the GW approximation\cite{aryasetiawan1998gw}), a material that was computed to be a \ac{DGM} with DFT-PBE could actually result to be a \ac{QSHI}.
    It is worth highlighting that the opposite could also be possible, where a material that is estimated to be a \ac{QSHI} within DFT-PBE could end up being a trivial insulator in a GW calculation (as discussed for the case of TiNI in Ref. \cite{marrazzo2019relative}).
    Second, a \ac{DGM} could be driven into a \zz topological state through an external perturbation that could open a gap in the material without driving a band inversion\cite{collins2018electric,zhao2020strain}.
    To this end, it is important to look at both the \ac{IG} and \ac{DG} of the material.
    The former is related to how strong a perturbation needs to be in order to drive the material in a semiconducting state, while the latter gives us an idea of how likely it is that a gap could be opened, without causing band inversion, or driving the material into a metallic state.
    Indeed, the higher the \ac{DG} of a material, the more difficult it will be to close it.
    For this reason, \acp{DGM} with small negative \ac{IG} and large \ac{DG} are prime candidates for further studies;  a visual summary of the direct and indirect gap of the \acp{DGM} is shown in Fig.~\ref{fig:dmg_hyst}.
    Among the various possible perturbations that could open the gap, the most interesting ones, from the standpoint of applications in a device, would be gating/electric fields\cite{Drummond.2012}, the interaction with a substrate\cite{menshchikova2013band} and/or the application of strain\cite{liu2014tuning}.

    For all the materials found, we searched the current literature to determine which \acp{QSHI} were already known, and which one are novel to this study.
    We find that among the 4 \acp{QSHI} identified, 3 were already known, namely the two tellurides of hafnium/zirconium, and hafnium bromide.
    HfBr\cite{zhou2015new,hirayama2018electrides} was already known both for being an easily exfoliable material and for its topological properties in its \ac{2D} and \ac{3D} forms.
    This material has a \ac{DG} of 233~meV and an \ac{IG} of 48~meV making it a good candidate for experiments and possible device realization.
    It should also be noted that this material belongs to a broader class of transition-metal halides MX (M=Zr,~Hf X=Cl,~Br) which has already been predicted to host topological properties.
    Furthermore, several class of similar honeycomb materials have also been predicted to host topological states such as transition metal carbides MC (M=Ti, Zr, Hf)\cite{zhou2016prediction} and transition metal compounds MM' (M=Ti, Zr, Hf) (M'=Bi, Sb)\cite{huang2018prediction}.

    HfTe$_{5}$ (ZrTe$_{5}$)\cite{weng2014transition} are two pentatellurides both of which have large \acp{DG} of 300(281) meV and \acp{IG} of 171(220) meV making them both very good candidates for \acp{QSHI} applications at room temperature.
    Interestingly, as shown by Weng et al.\cite{weng2014transition}, the presence of a band inversion in these materials comes mainly from the non-symmorphic features of their space group which, together with the presence of 2 non-equivalent Te atom chains, drives the ordering of the bands at $\Gamma$. 
    At the same time, the symmetries of the space group give raise to four-fold degenerate bands at the \ac{TRIM} points, 2 with even and 2 with odd parity, such that the presence of \ac{SOC} will only open a gap in a band structure that would otherwise be metallic, but will preserve the product of parity eigenvalues at the \ac{TRIM} points.
    These pentatellurides have also been proposed for the realization of the quantum anomalous Hall effect, when the transition metal Hf/Zr is substituted for a rare earth metal which could induce a magnetic ordering in the material, breaking time-reversal symmetry\cite{lowhorn2006enhancement,weng2014transition}.

    HgNS is a new compound which has not been discussed in the literature for its topological properties.
    Its direct and indirect gaps are both of 7~meV.
    Such low gap, in between that of silicene (1.4~meV) and germanene (23.8~meV) precludes room temperature applications.
    Nevertheless, at the gap, situated along the $\Gamma - Y$ direction, the material shows a linear band dispersion akin to that of germanene at $K$ when \ac{SOC} is taken into account.
    The similarity between the two systems, opens the possibility for a multitude of studies, such as the effect of an electric field on the gap and band inversion\cite{Drummond.2012,grassano2018detection} or the possibilities for opto-electronic applications in THz devices\cite{o2012stable}.
    Given that HgNS would constitute a new \ac{QSHI}, we also study it using the hybrid HSE06 functional as described in the methods.
    We find, as expected, that the effect of the hybrid is that of increasing the gap in the material.
    We can also observe from the band structure shown in the supplementary material\cite{supplementary} that the gap opening goes against the band inversion, moving the quasi-linear crossing toward the edge of the zone, but is not strong enough to drive the material to a trivial states.
    Hence, we find that HgNS is still a \ac{QSHI}, as corroborated by the \acp{HWCC} evolution and edge state calculations shown in the supplementary material\cite{supplementary}, even when a hybrid functional is used, and the gap increases from 7~meV to 38~meV.

    Among the \acp{DGM}, the most promising ones are Fe$_2$O$_3$ and Mo$_2$Ta$_2$O$_{17}$, with small \acp{IG} of -21 and -36 meV respectively.
    The small negative \acp{IG} indicate that it could be feasible to force those from being \ac{DGM} to a \ac{QSHI} by acting via strain, electric field or other perturbations.
    There is also HfCl which, like HfBr, is a transition metal halide and has been studied in Ref \onlinecite{zhou2015new}, in which the material is identified as a semi-metal for both PBE and PBE-SOC calculations, but as \ac{QSHI} in a HSE06-SOC calculation.
    This reiterates that \acp{DGM} identified by \ac{DFT} can or could actually be found to be \acp{QSHI} if studied with techniques, such as many body perturbation theory, that can correctly estimate the gap, or when exposed to a perturbation.

    \begin{table*}[h!]
      \caption{
        Table containing a collection of direct (\ac{DG}) and indirect (\ac{IG}) band gaps (derived from the final calculations on a dense k-point grid), binding energies ($E_B$) and previous literature references for the \acp{QSHI} and \acp{DGM} identified.
        In light gray, those in the previous work by Marrazzo et al.\cite{marrazzo2019relative} (in absence of strain).
        For HgNS, both the PBE and HSE gap (in parenthesis) are reported.
        }
      \centering
      \begin{tabular}{|c|c|c|c|c|c|}
        \hline
        & Formula   & \ac{DG} (meV) & \ac{IG} (meV) & $E_B$ (meV) & Ref. \\
        \hline
        \multirow{13}{*}{QSHI}
          & HfBr                                     &    233 &     48 &  15.8 & \cite{zhou2015new,hirayama2018electrides} \\
          & HfTe$_{5}$                               &    299 &    171 &  16.5 & \cite{weng2014transition} \\
          & HgNS                                     &      7 (38) &      7 (38) &  19.4 &  \\
          & ZrTe$_{5}$                               &    281 &    220 &  19.4 & \cite{weng2014transition} \\
          & \textcolor{gray}{Pt$_{2}$HgSe$_{3}$                      } & \textcolor{gray}{   178} & \textcolor{gray}{   142} & \textcolor{gray}{ 60.2} & \cite{marrazzo2018prediction} \\
          & \textcolor{gray}{RhTaTe$_{4}$                            } & \textcolor{gray}{    67} & \textcolor{gray}{    67} & \textcolor{gray}{ 26.4} & \cite{liu2017van} \\
          & \textcolor{gray}{AsCuLi$_{2}$                            } & \textcolor{gray}{    65} & \textcolor{gray}{    58} & \textcolor{gray}{ 62.7} &  \\
          & \textcolor{gray}{TiCu$_{2}$Te$_{3}$                      } & \textcolor{gray}{     6} & \textcolor{gray}{     5} & \textcolor{gray}{ 44.0} &  \\
          & \textcolor{gray}{IrNbTe$_{4}$                            } & \textcolor{gray}{    40} & \textcolor{gray}{    37} & \textcolor{gray}{ 27.1} & \cite{liu2017van} \\
          & \textcolor{gray}{Bi                                      } & \textcolor{gray}{   656} & \textcolor{gray}{   578} & \textcolor{gray}{ 17.9} & \cite{mounet2018two} \\
          & \textcolor{gray}{TiNI                                    } & \textcolor{gray}{    20} & \textcolor{gray}{    20} & \textcolor{gray}{ 14.6} & \cite{mounet2018two,wang2016band} \\
          & \textcolor{gray}{In$_{2}$ZnS$_{4}$                       } & \textcolor{gray}{     7} & \textcolor{gray}{     5} & \textcolor{gray}{ 36.1} &  \\
          & \textcolor{gray}{TaIrTe$_{4}$                            } & \textcolor{gray}{    15} & \textcolor{gray}{    15} & \textcolor{gray}{ 25.8} & \cite{liu2017van} \\
        \hline
        \multirow{11}{*}{DGM}
          & BaCr$_{2}$N$_{2}$O$_{8}$                 &     19 &   -124 &  11.0 &  \\
          & VBr$_{3}$                                &     33 &    -99 &  15.1 &  \\
          & FePO$_{4}$(CH$_{3}$)                     &     11 &    -58 &  18.8 &  \\
          & HfCl                                     &    264 &   -166 &  14.8 & \cite{zhou2015new} \\
          & Fe$_{2}$O$_{3}$                          &     17 &    -21 &   6.4 &  \\
          & MoI$_{2}$S$_{2}$                         &     47 &   -117 &  14.4 &  \\
          & Mo$_{2}$Ta$_{2}$O$_{17}$                 &     17 &    -36 &  10.0 &  \\
          & \textcolor{gray}{WTe$_{2}$                               } & \textcolor{gray}{   738} & \textcolor{gray}{   738} & \textcolor{gray}{ 24.7} & \cite{qian2014quantum,mounet2018two} \\
          & \textcolor{gray}{MoTe$_{2}$                              } & \textcolor{gray}{   107} & \textcolor{gray}{  -259} & \textcolor{gray}{ 24.5} & \cite{qian2014quantum,mounet2018two} \\
          & \textcolor{gray}{ZrBr                                    } & \textcolor{gray}{    45} & \textcolor{gray}{   -19} & \textcolor{gray}{ 15.6} & \cite{li2018high,zhou2015new} \\
          & \textcolor{gray}{ZrCl                                    } & \textcolor{gray}{    64} & \textcolor{gray}{  -196} & \textcolor{gray}{ 14.7} & \cite{zhou2015new} \\
        \hline
      \end{tabular}
      \label{table:results}
    \end{table*}

    When considering the \acp{QSHI} identified in this work, we obtain a relative abundance of topological insulators of 0.5\%, which goes up to 1.4\% if the \acp{DGM} are included as well.
    This result is comparable to that of the previous work of Marrazzo et al\cite{marrazzo2019relative} where an abundance of 1.1\% was found for \acp{QSHI}.
    The difference in the values can be explained by two reasons: first, in the work from Marrazzo, strained configurations had also been considered when determining a material topology.
    Second, when analyzing the space group distributions of the two datasets, shown in Fig.~SM1-2, we can observe that the old dataset contained more materials with monoclinic Bravais lattice, in particular with space group 11, which from the analysis seems to be the one with the most \acp{QSHI} and \acp{DGM} when compared to all other space groups.

  \section{Conclusions}
    We performed a systematic high-throughput study of \ac{2D} exfoliable materials in order identify novel \acp{QSHI}.
    Starting from a set of 783 materials we found 4 \acp{QSHI} (1 novel) and 7 \acp{DGM} (6 novel) that could be driven into a \ac{QSHI} state by material engineering, for a relative abundance of 0.5\% for \acp{QSHI} and 1.4\% if \acp{DGM} are also included.
    Of the 4 \acp{QSHI} identified, the 3 with the largest gaps had already been explored in the literature.
    In particular, the two pentatellurides HfTe$_{5}$ and ZrTe$_{5}$ exhibit both large direct and indirect gaps which makes them promising candidates for room-temperature applications.
    The remaining material, namely HgNS, represents a newly identified \ac{QSHI} with linear bands in proximity of its direct gap of 38~meV when computed with an hybrid HSE functional.
    The main feature of this material is the small gap and band linearity, akin to that of silicene or germanene.
    For this reason, HgNS could be a promising candidate for future electronic and photonic applications in the low energy regime.

    All materials identified in this work are easily exfoliable, with a binding energy lower than 20 meV/\angstrom$^2$\cite{mounet2018two} and for this reason they are prime candidates for possible experimental realization through exfoliation techniques starting from their experimentally known 3D bulk form.

  \section{Acknowledgments}
    This work was supported by the National Centre for Computational Design and Discovery on Novel Materials (NCCR MARVEL) of the Swiss National Science Foundation.
    D.G.  gratefully acknowledge support from the EU Centre of Excellence, MaX Materials design at the eXascale (Grant 824143).
    We acknowledge PRACE for awarding us access to Marconi at Cineca, Italy (project id. 2016163963).
    A.M. acknowledges financial support from ICSC - Centro Nazionale di Ricerca in High Performance Computing, Big Data and Quantum Computing, funded by European Union - NextGenerationEU
  \bibliography{bibliography}

\end{document}